\documentclass[10pt,a4paper,floatsintext]{article}
\usepackage{geometry}
\usepackage[utf8]{inputenc}
\usepackage{amsmath}
\usepackage{amsfonts}
\usepackage{amssymb}

\usepackage{caption}
\usepackage{threeparttable}
\usepackage{booktabs}
\usepackage{listings} % for list of JATSdecoder example output
\usepackage[T1]{fontenc} % for < > signs
\usepackage{multicol}
\usepackage{multirow}
\usepackage{adjustbox}
\usepackage{stackengine}

\usepackage{lscape}

\usepackage{comment}

\usepackage{url}
\usepackage{hyperref}
\usepackage{apacite}
\usepackage{tikz}
\usetikzlibrary{positioning, shapes, arrows.meta, fit}

\title{Extraction of tabulated statistical results with tableParser}
\author{Ingmar Böschen; ORCID: 0000-0003-1159-3991}

\begin{document}
\maketitle

\begin{center}
Abstract
\end{center}

Tabulated content is omnipresent in scientific literature. 
In this work I describe the functionality of the R package \textit{tableParser}, which is designed to extract and postprocess tables from NISO-JATS-encoded XML, any other HTML, DOCX, and, with some restrictions, PDF documents. 
\textit{tableParser} was developed in order to extract and analyze tabled statistical test results reported in scientific publications. 
On a big scale, \textit{tableParser} can be used to analyze the extent and variation of effects of specific variables on others, reporting practices, or summarization of results. 
On a single document scale, the automated extraction of statistical test results can also be beneficial for completeness and consistency checkups of tabulated test results in unpublished documents. 

Documents can be processed in three different decoding depths. 
On a first level, all tables within a document are compiled into a list of character matrices with captions and footnotes as attributes with \textit{table2matrix()}. 
The matrix content can be collapsed into a human-readable text string by imitating the behavior of a screen reader with \textit{table2text()}. 
Optionally, and to reduce cross references, many common codings that are reported within the table's caption and footnote can be used to decode and expand the table's content. 
Finally, the collapsed and decoded table content can be further processed to best match an ideal input for the extraction of statistical standard results with the \textit{standardStats()} function from the \textit{JATSdecoder} package. 
Finally, the output of \textit{table2stats()} is a data frame with all detected standard results as columns, and, if calculation is possible, a recalculated p-value. 
If desired, an automated consistency check of the reported and the coded p-values with the recalculated p-value can be initiated. 

In general, \textit{tableParser} works best on rather simple, but also some more complex, barrier-free HTML tables that are encoded in the NISO-JATS standard, where captions and footnotes are clearly identifiable. 
By guessing the tables captions and footnotes conservatively, the processing of tables within any other HTML and DOCX documents is comparably robust. 
Technically, tables within PDF documents often fail to be extracted correctly, and captions and footnotes cannot be detected with the implemented functions. 
Therefore, a precise decoding of codes within a table is not possible for PDF input which lowers its detection accuracy of statistical results.

\section{Objective}
\label{sec:intro}
Extracting interpretable information from tables is a difficult task that, despite the impressive capabilities of modern large language models, after 30 years of research, is still far from being solved completely \cite{shigarov2023}. 
Due to the great technical variety in structures and formats, decoding tabled information is a research field that has been considered by many research communities under differing perspectives \cite{bonfitto2021table}. 
The automated processing and/or conversion of effective tables to other formats, principally aimed at storing and transmitting structured information, are still open problems, which fall under the umbrella of \textit{Table Understanding problems} \cite{bonfitto2021table}.

Rather than raw data tables, effective tables are designed for being interpreted by humans. 
They represent a very dense and sensible way of reporting descriptives, showing up differences between groups, associations between variables by statistical tests, and summarizing model results. 
Statistical result tables often contain encoded information (e.g., superscripted letters, abbreviations, special signs), and, in addition, graphical features, such as lines, grouped or colored cells, as well as different text styles (bold, italic) to encode the presented results effectively. 
In practice, result tables often contain a lot more information when compared to the textual description of the tabled results, which focuses on the main or most interesting results. 

Depending on the complexity and structure of a table, reading out the full information and converting it into language-based information is a challenging task. 
When reading out statistical result tables, one needs to navigate within the table's row and column namings and groupings, have knowledge about what is reported exactly inside the cells, the meaning of bracketed numbers, and often decode abbreviations and other codings that are reported in the table footnotes (e.g., `*' for p<.05). 
The decoding complexity is compounded by empty cells that may be used for graphical grouping or as an indication of a nonexistent result. 
Furthermore, and for example, in the case of encoded p-values, the absence of a code constitutes information. 

Taking all these obstacles into account, the R \cite{RCore} package \textit{tableParser} \cite{tableParser} imitates the functionality of a screen reader for blind people that is specialized in handling statistical result tables. 
The main goal leading the development is to extract and unify tabled statistical standard results reported in the scientific literature. 
The package contains a function collection to extract tables from documents, transform the tabled content into a text vector, and an easy-to-process data frame with the detected statistical standard results. 
Solutions for standardizing the textual representations of hexadecimal and HTML encoded special characters, as well as the textual representations of numbers and the correction of commas, are implemented. 
Before collapsing the tabulated content, several decoding and expansion processes can be initiated to reduce cross-references with the text of captions and footnotes. 
The heuristic-driven processes are solely based on an iteratively developed set of regular expressions. 
Still, \textit{tableParser}'s functionality is very general and can be applied to any tabulated input with individual settings.

\subsection{Technical representations of tables in documents}
Following the \textit{three-part-table} approach, here, the term table is used for matrix-like structures with an accompanying caption and/or footnote. 
In practice, at least two common technical approaches are used to encode tables in HTML documents that may result in the exact same visual appearance. 
The three parts of a table may be encoded in a floating stream of paragraphs (<div>, and/or <p> tags) with individual graphical settings. 
UI-based text editors will often create floating paragraphs to display a three-part-table.
The Journal Article Tag Suite (NISO-JATS) \cite{JATS} is a well-established HTML markup standard to store and share scientific content on the web. 
Table structures are stored within a <table-wrap> tag. 
Besides the main matrix that is encoded within the <table> tag, a <caption> and <table-wrap-foot> tag may be appended to the structure. 

The wrapping of the three parts of a table in NISO-JATS facilitates the association of the tabled content to the surrounding information.
As tables may also be reported without a caption and/or footnote, the surrounding tags of a table in the floating paragraph style may contain the surrounding blocks of text of the manuscript. 
Further, multiple paragraphs may be used to display multi-row captions or footnotes. 
This complicates a precise extraction of three-part-table structures in non-NISO-JATS-encoded documents.

\begin{figure}[htbp]
\caption{Three-part-table structures with floating paragraphs and NISO-JATS standard}
\centering

\begin{minipage}{.4\textwidth}
\centering
\caption*{\textbf{floating paragraphs}}
\begin{tabular}{|l|}
\hline
<div>\\
\hspace{.2cm}  <p>Caption</p>\\
</div>\\
<div>\\
\hspace{.2cm}   <table>...</table>
</div>\\
<div>\\
\hspace{.2cm}  <p>Footnotes</p>\\
</div>\\
\hline
\end{tabular}
\end{minipage}%
\begin{minipage}{.4\textwidth}
\centering
\caption*{\textbf{NISO-JATS standard}}
\begin{tabular}{|l|}
\hline
<table-wrap>\\
\hspace{.2cm}   <caption>...</caption>\\
\hspace{.2cm}   <table>...</table>\\
\hspace{.2cm}   <table-wrap-foot>...</table-wrap-foot>\\
</table-wrap>\\
\hline
\end{tabular}

\end{minipage}%

\label{fig:HTMLcodings}
\end{figure}

Tables within DOCX documents are structured by the floating paragraph approach (with <w:p>-tags). 
The DOCX tag equivalent to the HTML <table>-tag is <w:tbl>. 
The raw coded content can be accessed from within the XML file that is available after unpacking a DOCX file.

Since tables within PDF documents are not coded in a well-structured manner, it is comparatively hard to reliably extract them. 
Mostly, they are implemented in a purely graphical way, without document sections or table references. 
The R package \textit{tabulapdf} \cite{tabulapdf} gives quick access to the Tabula Java library. 
Its function \textit{extract\_tables()} is implemented in \textit{tableParser} to extract tables from text-based PDF files. 
Since it is not possible to extract the table captions and footnotes from PDF documents with \textit{tabulapdf}, and the table extraction often goes along with spurious artifacts, in practice, the processing of tables in PDF documents is less precise.

\section{The \textit{tableParser} workflow and main modules}

\textit{tableParser} contains three convenient functions to extract and collapse tables. 
They can be applied to NISO-JATS-encoded XML, any other HTML, DOCX, and PDF files. 
The function \textit{table2matrix()} extracts the tables within a document and returns a list of character matrices with the tabled content. 
The detected table caption and footnotes, as well as the result of a table classifier, are appended as attributes to each matrix. 
With several user-definable uniformizations and decoding depths, the matrix content can be collapsed into a text string with \textit{table2text()}. 
The collapsed and decoded matrix can be processed and screened by the \textit{table2stats()} function. 
It makes use of the \textit{standardStats()} function from the \textit{JATSdecoder} package \cite{JATSdecoderCRAN1.2} in order to extract and process the statistical standard test results presented in a table. 

\begin{table}[htbp]
\caption{Main modules of \textit{tableParser} that can be applied on NISO-JATS-encoded XML, any other HTML, DOCX, and PDF documents}
\centering
\begin{tabular}{ll}
\toprule
Function & Output\\
\midrule
\textit{table2matrix()} & tabled content as list of matrices\\
\textit{table2text()} & collapsed and (partyly) decoded tabulated content as text vector\\
\textit{table2stats()} & unified statistical standard results as data frame object\\
\midrule
\end{tabular}
\label{tab:mainModules}
\end{table}

\begin{comment}
A simplified visualisation of the principal workflow that is performed to extract, decode and collapse tabled content into a text string, is displayed in Figure \ref{fig:simpleFlow}. \\

\begin{figure}[htbp]
\caption{Simplified process chart of the table to text to statistics conversion}
    \centering 
    \resizebox{\textwidth}{!}{%
\begin{tikzpicture}[
    % Styles
    node/.style={
        draw=blue!60, 
        fill=blue!10,
        rectangle,
        rounded corners,
        font=\sffamily,
        minimum width=1.5cm,
        minimum height=1cm,
        text centered,
        inner sep=5pt
    },
    edge/.style={
        draw=black!80,
        ->,
        thick,
        line width=1.2pt,
            >={Stealth[length=4mm, width=2mm]} % Use Stealth arrow tips for elegance
    },
    every node/.style={inner sep=0pt} % Remove unnecessary padding
]
    % Knoten
    \node[node] (a) {document};
    \node[node, right=of a] (b) {table structure};
    \node[node, right=of b] (c) {character matrix};
    \node[node, right=of c] (d) {collapsed text};
    \node[node, right=of d] (e) {statistics};
 %    \node[node, right=of b] (c) {classifier};
    \node[node, below=of c] (f) {caption/footnote};

    % Kanten
    \draw[edge] (a) -- (b);
    \draw[edge] (b) -- (c);
    \draw[edge] (c) -- (d);
    \draw[edge] (d) -- (e);
    \draw[edge] (b) -- (f);
    \draw[edge] (f) -- (d); 
%    \draw[edge] (f) -- (c); 
    \end{tikzpicture}
    }%
\label{fig:simpleFlow}

\end{figure}
\end{comment}

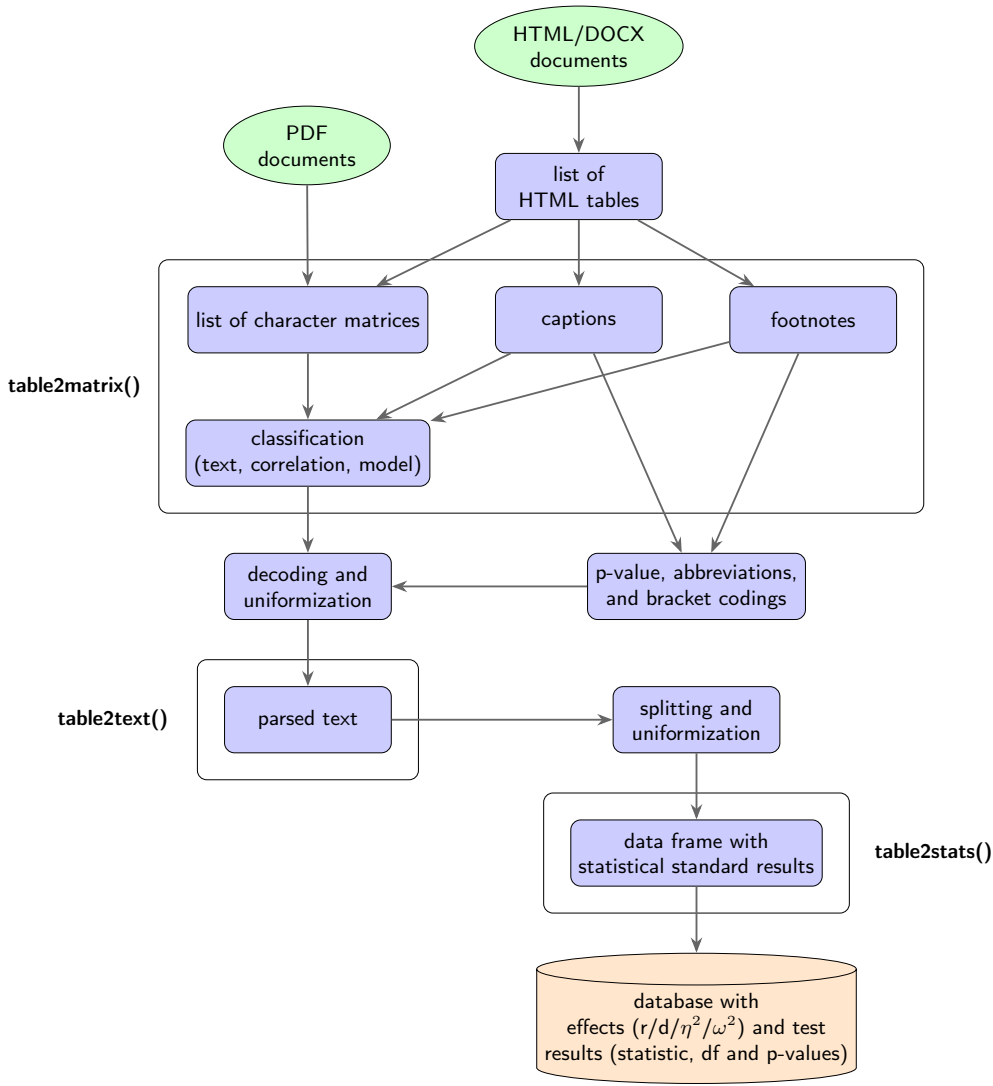
\begin{figure}[htbp]
 \caption{File format-specific processing of documents to create a database with statistical results}

    \centering 
    \resizebox{.8\textwidth}{!}{%
\begin{tikzpicture}[
   % node distance=1.5cm and 2cm,
    every node/.style={rectangle, draw, minimum width=2.5cm, minimum height=1cm, font=\sffamily\small},
    startstop/.style={ellipse, fill=green!20},
    process/.style={rectangle, fill=blue!20, rounded corners=0.15cm},
    result/.style={rectangle, fill=orange!20, rounded corners=0.0cm},
    decision/.style={diamond, fill=yellow!20, aspect=2, shape border rotate=45, inner sep=0},
    empty/.style={rectangle, fill=white, color=white},
    database/.style={cylinder, shape border rotate=90, aspect=0.1, fill=orange!20, minimum height=1.5cm},
    arrow/.style={thick, -Stealth, draw=black!60},
    myfit/.style={draw, solid, rounded corners, inner sep=0.4cm, label={left:table2stats()}},
    myfit2/.style={draw, solid, rounded corners, inner sep=0.4cm, label={left:\textbf{table2matrix()}}},
    myfit3/.style={draw, solid, rounded corners, inner sep=0.4cm, label={left:\textbf{table2text()}}},
    myfit4/.style={draw, solid, rounded corners, inner sep=0.4cm, label={right:\textbf{table2stats()}}},
]

%%%%%%%%%%%
% Nodes
\node (start) [startstop, align=center] {HTML/DOCX\\ documents};

\node (table) [process, below=of start, align=center] {list of\\ HTML tables};

\node (startPDF) [startstop, align=center, left=of start, xshift=-.25cm, yshift=-1.5cm] {PDF\\ documents};

\node (matrix) [process, below left=of table] {list of character matrices};
\node (caption) [process, below=of table] {captions};
\node (footer) [process, below right=of table] {footnotes};

\node (legend) [empty, below=of footer, xshift=-1.75cm, yshift=.0cm, align=center]{};
\node (legend2) [process, below =of legend, align=center, yshift=.0cm] {p-value, abbreviations, \\ and bracket codings};

\node (classify) [process, below=of matrix, align=center] {classification \\
(text, correlation, model)};

\node (decode) [process, below=of classify, align=center] {decoding and\\
uniformization};

\node (text) [process, below=of decode] {parsed text};

\node (split) [process, below=of legend2, align=center, yshift=-.0cm] {splitting and\\ uniformization};
\node (stats) [process, below=of split, align=center] {data frame with \\ statistical standard results};

\node (data) [database, below=of stats, align=center] {database with\\ effects (r/d/$\eta^2$/$\omega^2$) and test\\ results (statistic, df and p-values)};

%%%%%%%%%%%%%%%%
% Ractangles around nodes
%\node [myfit] (fitbox) [fit = (table) (classify) (text) (footer)] {};
\node [myfit2] (fitbox) [fit =  (matrix) (caption) (footer) (classify)] {};
\node [myfit3] (fitbox) [fit =  (text)] {};
\node [myfit4] (fitbox) [fit = (stats)] {};

%%%%%%%%%%%%
% Arrows
\draw[arrow] (start) -- (table);
\draw[arrow] (startPDF) -- (matrix);

\draw[arrow] (table) -- (matrix);
\draw[arrow] (table) -- (caption);
\draw[arrow] (table) -- (footer);

\draw[arrow] (matrix) -- (classify);
\draw[arrow] (caption) -- (legend2);
\draw[arrow] (footer) -- (legend2);
\draw[arrow] (caption) -- (classify);
\draw[arrow] (footer) -- (classify);

\draw[arrow] (classify) -- (decode);
\draw[arrow] (decode) -- (text);
%\draw[arrow] (legend2) -- (classify);
\draw[arrow] (legend2) -- (decode);
\draw[arrow] (text) -- (split); %node[midway,above] {standardStats()};
\draw[arrow] (split) -- (stats);
\draw[arrow] (stats) -- (data);

\end{tikzpicture}
}%
\label{fig:flowFull}

\end{figure}

Figure \ref{fig:flowFull} displays the file format-specific processing of tables and outputs (boxes) of the three main modules of \textit{tableParser}. 
Data frames generated with \textit{table2stats()} can be easily merged and further analyzed to compile a large-scale database containing the extracted tabulated statistical standard results from a literature collection.
Since the extraction of table captions and footnotes is not possible for PDF documents, table processing is less sophisticated for this file format. 

\subsection{Conversion of HTML tables into a list of matrices}
The function \textit{table2matrix()} contains file-specific table-to-matrix conversion solutions for NISO-JATS-encoded XML, HTML, DOCX, and PDF format. 
The routines contained in the function are also available individually. 
\textit{get.HTML.tables()} extracts HTML-encoded table structures within NISO-JATS-encoded XML and HTML files and returns a vector of plain HTML tables. 
For three-part-table structures encoded with the <table-wrap>-tag, the caption of each table is extracted with \textit{get.caption()}, and the footnotes with \textit{get.footer()}. 
The matrix part of the HTML-table structure can be converted into a two-dimensional character matrix with \textit{table2matrix()}. 

Tables within DOCX documents can be extracted and returned as a list of character matrices with the function \textit{docx2matrix()}. 
It unpacks the DOCX file to a temporary folder and extracts the DOCX-specific encoded tables from within the document's raw XML file. 
The function \textit{guessCaptionFooter()} is designed to extract the caption and footnote in documents with floating <div>-tag encoded three-part-table structures. 
In standard mode, it extracts the surrounding non-empty <div>-tags around a table. 
Preceding tags that have only one sentence are interpreted as caption. 
A subsequent tag with less than six sentences is returned as footnote text. 
Tables within PDF files are extracted with the \textit{extract\_ tables()} function from the \textit{tabulapdf} \cite{tabulapdf} package. 
Text and table extraction from PDF files often results in errors. 
Tests on input from different publishers showed that the results of the PDF-to-matrix conversion are often messy and have to be treated with care. 

Information within HTML tables often contains further HTML and/or hexadecimal-encoded elements, such as special letters, stylings, hyperreferences, or formulas. 
To convert and unify hexadecimal-encoded letters to a Unicode format, \textit{JATSdecoder}'s function \textit{letter.convert()} can be activated in \textit{table2matrix()}. 
The option `\textit{greek2text}' enables the conversion of common Greek letters into a Latin text format ($\alpha$ -> alpha, $\beta$ -> beta). 
Activating the option `rm.html' will remove all further HTML tags but replace the opening superscript tags with the hat sign (`\string^') and the subscript tags with a low line (`\_').
Further, the option `\textit{collapseHeader}' can be activated to collapse multiple connected header rows into a single row. 

For HTML and DOCX files, the \textit{table2matrix()} and \textit{docx2matrix()} options `\textit{replicate}` and `\textit{repNum}' control the handling of content within connected cells. 
In the standard configuration, the content of a connected cell is only inserted into the first of the disconnected cells. 
The content is copied into each of the disconnected cells when the argument `\textit{replicate}' is activated. 
To omit the replication of numerical results (any letter-number combination pointing to a number with an operator) reported in connected cells, the argument `\textit{repNums}' is deactivated in standard mode.

\subsection{Table types}
\label{sec:classifier}

Since different types of tables demand individual handling, the extracted matrix objects are classified based on the tabulated content and references within the caption and footnote. 
Up to now, the function \textit{tableClass()} classifies tables into one of a limited number of classes (vector, text, tabled result, matrix, correlation matrix, model/multi-model with model statistic). 
The table class is attached to the tables extracted with \textit{table2matrix()} as attribue. 

Matrices with only one row or one column are tagged as `vector'. 
If a matrix contains more than 80\% of its nonempty cells starting with a letter, it is classified as `text'.
In general, matrices with mostly numeric cell content are classified as `tabled result'. 
However, the tabled results may be further specified. 
Tables with a sequence of increasing numbers in their first row and first column and tables with at least three matching cell labels in the first row and first column are tagged as `matrix'. 
Correlation matrices are a special case of a `matrix' type table.
They are detected by further screening the table's caption for search terms like `correlation', `association', and `reliability'. 
A check of the cells plausibility to be a correlation (values in [-1,1]) and a search for increasing numbers in the row and column names also allow the detection of correlation matrices when no table caption is applicable. 
If a table represents a mixture of a tabled result and a correlation matrix, the returned class is `correlation', but descriptives and correlations will be treated separately during conversion. 
Tables that contain only one numerical entry in lines with model statistics (e.g., $R^2$, AIC, BIC) and contain the term `regression' or `model' in their caption are classified as `model with model statistic'. 
If more than one numerical value of any model statistic is reported, `multi-model with model statistic' is returned.

As some decisions of the classifier depend on an applicable table caption or may not be adequate, a possibility to override its decision for table collapsing is implemented in \textit{table2text()} with the argument `\textit{forceClass}'.

\subsection{Decompression of codings in tables}
In scientific publications, result tables often contain special signs, superscripted letters, abbreviations, and bracketed numbers that can be decoded only with an adequate decoding scheme, mostly supplied in the table footnotes. 
In many cases, an elegant decoding and special handling of codes is essential for a precise extraction of statistical results and consistency checks with \textit{JATSdecoder}'s function \textit{standardStats()}.
The aim of \textit{tableParser}'s table-to-text conversion is to decode as much of these references and codings as possible. 
A particular focus is on decoding statistical standard results.

The \textit{tableParser} function \textit{legendCodings()} is built to detect codings and abbreviations in footnotes. 
It returns a list of vectors with specific codings and references that is used at different stages of the table-to-text conversion process implemented in \textit{table2text()} and \textit{table2stats()}. 
The decoding depth and standards can be controlled by the user.

\subsubsection{p-values}
The most often calculated and most often encoded result in statistical result tables is the p-value. 
It describes the probability of obtaining the found or even more extreme results, given the postulated null hypothesis (and assumptions) is true, and serves as a decision criterion for rejecting null hypotheses. 
Any p-value below a pre-specified threshold (the alpha level $\alpha$) is considered to be significant. 
The value of $\alpha$ is an a priori set probability to falsely reject a tested nullhypothesis. 
In practice, it is quite common to work with an $\alpha$ level of 5\% \cite{study.character} and encode p-values with (superscripted) special characters.
Also, bold and italic text styling is used to encode p-values below or above the $\alpha$ level ('significant effects in bold').

By activating the argument `\textit{decodeP}', the p-value codings within the table are replaced by a text string. 
To clearly distinguish between coded and not coded p-values for later processing, a special operator (';;') is used in front of the imputed p-value. 

The most common standard is to encode p<.05 with '*', p<.01 with '**' and p<.001 with '***'. 
This scheme can be set as standard in the table-to-text conversion process with the argument `\textit{standardPcoding}'. 
However, if a deviating coding scheme is recognized by \textit{legendCodings()}, the standard scheme is not applied.

A non-trivial problem arises when only significant p-values are coded. 
Technically, the absence of a coding is a reference to a nonsignificant value ($p>\alpha$), and therefore, often a checkable detail of a result as well. 
To impute nonsignificant p-values, the option \textit{noSign2p} can be activated. 
In standard result tables, values without a p-coding in columns with p-coding are considered to be reported as not significant. 
The highest detected p-value code ($\alpha$) is extracted and attached to the cell's content with the easy-to-identify separator (';; p>numeric $\alpha$-value').
In correlation matrices with codes for significant correlations, to all cells that were classified as correlations and that do not have a code for a p-value, ';; p>numeric $\alpha$-value' is attached.

It must be noted that a comparably high rate of false positive non-significant imputations has to be expected in highly complex table structures. 
In columns or correlation matrices that do not have at least one result coded as significant, no imputation is performed. 
This may result in missing nonsignificant p-value imputations. 

\subsubsection{Numeric values in brackets}
Sometimes, two different kinds of numeric results are presented within the same cell of a table. 
The meaning of the bracketed value is either reported in the table header, or within the footnote. 
These two representations are treated differently before collapsing the table to text. 
If the header line contains the reference for the values in brackets, a new column is created with the name and numeric values reported in brackets, just without brackets. 
For bracketed values that represent percentages, the brackets around the values are removed, \%-signs are attached to the numeric value. 
If a reference to the meaning of bracketed numbers is found in the footnote, again, this reference is imputed in front of the values, just without brackets. 
Up to now, the extraction of meaning for bracketed numbers from the footnote is limited to standard errors and deviations, reliability alike coefficients alpha, confidence intervals and the square root average variance explained. 
Other statistics are neither recognized within the footnotes, nor imputed for the brackets.

\subsubsection{Abbreviations}
As the width of a publishable table is often limited by publication requirements, abbreviations are used to reduce the size of a table. 
To enhance the identification of effects associated with specific variables, it is beneficial to expand the abbreviations into their full labels. 
\textit{legendCodings()} detects multiple variations of abbreviations. 
It extracts the capital letter code and the abbreviated label when the abbreviations are listed with the '=' or ':' sign or as dot abbreviated words followed by ',' or ';'. 
Also, abbreviations reported in brackets can be detected. 
Activating the option \textit{expandAbbreviations}, enables the decoding of detected abbreviations to their full names in the table-to-text conversion process. 
This will enhance readability of the output and will not affect the extraction of statistical standard results. 

\subsubsection{Superscripts}
In general, and besides for p-values, superscripted characters are used to add supporting textual information to a result. 
If the option \textit{superscript2bracket} is activated, superscripted characters, numbers and special signs in a cell are replaced by the label extracted from the footnote.
The textual content of the superscript is presented within brackets. 
Again, this will improve readability by decreasing the number of cross-references to the footnote. 
Notably, superscripted codes within a table that are reported without a superscript in the footnote cannot be handled. 

\subsubsection{Sample size/degrees of freedom} 
Some statistical test results are only checkable when they are reported with degrees of freedom (df).
For the two sample t-test and the bivariate correlation coefficients, the sample size subtracted by 2 is a good estimate for the df of the reported test. 
Therefore, all numbers reported behind the letter `N' in the caption and footnote are extracted, interpreted as sample size, and can be imputed as df-value (N-2) to the parsed tabled content, by activating the argument \textit{dfHandling}. 
This is especially useful for correlation matrices that mostly do not contain the df for the reported tests, as well as for tables with t-values but no reported df-values. 
This option has to be handled with care, though. 
For tables displaying more than two parameter estimates and a t-value but no df, the imputation of N-2 overestimates the nominal df-value. 
Notably, this bias vanishes with increasing sample size, since t converges to a standard normal with increasing degrees of freedom. 
In correlation matrices, the imputation relies on the assumption of non-missing data. 
If pairwise complete observations are reported, N-2 overestimates the actual df. 
This results in an underestimated standard error and, therefore, invalidly recalculated p-values.

\subsection{Collapsing table structures to text}
In this section a description of the implemented table preprocessing, collapsing, and decoding rules is given. 
Most tabled statistical results in the literature are reported in a so called multi-way table structure.
%Therefore, \textit{tableParser} is especially designed for such tables. 
Multi-way tables consist of an outer part with the table row and column labels, and an inner part, that contains the specific results, whereas single way matrices have column labels only. 
A detailed overview on the problems arising when interpreting multi-way tables is given by \citeA{shigarov2023}. 
Raw data is mostly stored in one-way tables and comparatively easy to extract. 
To give a graphical comparison, Table \ref{tab:simpleTable} displays an overly simplified example of a one- and a multi-way table with numerical, character, and two distinct textual representations of a nonpresent/missing value. 

\begin{table}[htbp]
\caption{Example for a single and a multi-way table structure}
\centering

\begin{minipage}{.4\textwidth}
\caption*{\textbf{One-way table example}}
\centering
\begin{tabular}{|c|c|c|}
\hline
Column A & Column B & Column C\\
\hline
1 & 3  & A \\
\hline
2 & 4  & - \\
\hline
\end{tabular}
\end{minipage}%

\bigskip
\begin{minipage}{.4\textwidth}
\caption*{\textbf{Multi-way table example}}
\centering
\begin{tabular}{|l|c|c|}
\hline
 & Column A & Column B\\
\hline
Row A & 1  & A \\
\hline
Row B & 2  & NA \\
\hline
\end{tabular}
\end{minipage}%

\label{tab:simpleTable}
\end{table}

One-way tables are often used to store raw data with a row per element and a column per variable. They can be read out line- or column-wise. 
Multi-way tables are often used to present aggregates of the raw data, such as statistical results. 
These tables can quickly become complicated to read out. 
The most neutral information extraction approach is to consider each cell as an individual result of the corresponding row and column naming. 
Every cell of the inner matrix is expected to relate to a row and column label. 
Here, this read-out procedure will be referred to as matrix-wise extraction. 

Another convenient procedure is to read out the inner cells row- or column-wise. 
In the row-wise approach, the row label is considered to be a global description for all inner cells within the row. 
Only the column name is connected to the cell's content; the row name stands in front as a descriptor. 
The column-wise read-out procedure treats the column name as a descriptor and connects only the row names to the cell's content.

Table \ref{tab:extractionResults} illustrates the difference in the results of the three different read-out procedures based on the content of the multi-way example table in Table \ref{tab:simpleTable}. %, collapsed by the matrix-, row-, and column-wise approach.

\begin{table}[htbp]
\caption{Matrix-, row- and column-wise cell collapsing of the multi-way example matrix in Table \ref{tab:simpleTable}}
\small
\centering
\begin{threeparttable}
\begin{tabular}{lll}
\toprule
matrix-wise & row-wise & column-wise \\
\midrule

$\left[1\right]$ Row A, Column A=1 & $\left[1\right]$ Row A: Column A=1, Column B: A & $\left[1\right]$ Column A: Row A=1, Row B=2\\

$\left[2\right]$ Row A, Column B: A & $\left[2\right]$ Row B: Column A=2, Column B: NA & $\left[2\right]$ Column B: Row A: A, Row B: NA\\

$\left[3\right]$ Row B, Column A=2 & & \\
$\left[4\right]$ Row B, Column B: NA & & \\
\midrule
\end{tabular}
\begin{tablenotes}
\item[Numbers in squared brackets represent the position within the result vector.] 
\item[Note: Numeric content is collapsed with `=', textual content with `:'.]
\end{tablenotes}
\end{threeparttable}
\label{tab:extractionResults}
\end{table}

All three collapsing modes are implementent in the table collapsing function \textit{table2text()} and, for already available character matrices, in \textit{parseMatrixContent()}. 
In standard mode, the selection of the collapsing standard is based on the table classifier. 
The fallback readout procedure is row-wise. 
If a table is detected to be of type `matrix', the matrix-wise approach is used, with the special operator `$<<\sim>>$' to connect row and column name. 
Before collapsing, the term `VALUE' is attached to the numerical values of the matrix with the equal sign (e.g., `A $<<\sim>>$ B, VALUE=0.25'). 
In correlation matrices, the term `VALUE' is replaced with the letter `r'. 
Since values that are on the diagonal of a correlation matrix and that are not equal to 1 do not represent correlations, the `VALUE' reference is replaced with `alpha' or 'sqrtAVE', depending on the reference made in the caption or footnote. 

The option `forceClass' enables users to manually select the table class for collapsing (`tabled result', `matrix', `correlation', `text'). 
Switching from row-wise to column-wise extraction can be achieved by activating the option `rotate' or by transposing the matrix input manually.

\subsection{Processing of more complex table structures}
In practice, table structures are often a lot more complex. 
Tabled results may come with references and codings, which are described within the table itself or the footnote. 
To give insights into the fine tuned heuristics to unify and decompress complex tables into simple multi-way tables before collapsing, some practical examples of more complex table structures are given in this section. 
The examples will illustrate how the function \textit{table2text()} preprocesses and handles complex representations of tabled results and codes that are reported in the footnotes. 
Besides the original table, the preprocessed (unified and decompressed) matrix content is displayed. 
To illustrate the usefullness for the later extraction and postprocessing of statistical standard results with \textit{standardStats()}, table specific features are introduced. 
All example tables listed here are stored within an example DOCX document that is available in the Appendix or the \textit{tableParser} repository: \href{https://www.github.com/ingmarboeschen/tableParser}{https://www.github.com/ingmarboeschen/tableParser}).

\subsubsection{Example 1}
Compared to the introductory example, the first example table is just a bit more complex. 
Table \ref{tab:complexTable1} contains several empty cells within the row labels and numbers in brackets, that are connected to two distict statistics, coded in the header label. 
The empty cells within the first column are a visual helper to orientate within the table. 
Visually, it is not possible to conclude about the technical representation within the table. 
In fact, the cells may be empty. 
Also, it is possible, that all cells below and within the `Age' and `Gender' cell, are connected, with a top aligned text. 

\begin{table}[htbp]
\caption{Example of a more complex table displaying categorized descriptive meassures and the uniformization of empty row names and measures in brackets}

\begin{minipage}{1\linewidth}
\centering
\caption*{\textbf{Original table}}
\begin{tabular}{llcc}
\toprule
Variable & Categories & Freq. (\%) & Mean (SD) income \\
\midrule
Age & 20–25 & 78 (39.79) & 1,345 (250)\\
 & 26–31 & 65 (33.16) & 1,825 (535)\\
 & 32 and older & 53 (27.05)& 2,315 (930)\\
Gender & Men & 97 (49.49) & 1,965 (550)\\
 & Women & 99 (50.51) & 1,685 (350)\\
\midrule
\end{tabular}
\end{minipage}%

\bigskip
\begin{minipage}{1\textwidth}
\centering
\caption*{\textbf{Unified and preprocessed matrix}}
\begin{tabular}{|l|l|l|l|l|}
\hline
 & Category & Freq. & Mean income & SD\\
\hline
Variable: Age & 20-25 & 78, 39.79\% & 1345 & 250\\
\hline
Variable: Age & 26-31 & 65, 33.16\% & 1825 & 535\\
\hline
Variable: Age & 28 and older & 53, 27.05\%& 2315 & 730\\
\hline
Variable: Gender & Men & 97, 49.49\% & 1965 & 550\\
\hline
Variable: Gender & Women & 99, 50.51\% & 1685 & 350\\
\hline
\end{tabular}
\end{minipage}

\label{tab:complexTable1}
\end{table}

The unified and preprocessed matrix has a simple multi-way table structure. 
The empty cells in the first column now contain the value of the last non empty cell in that column. 
The content of the first cell was removed and added in front of every other cell in the first column. 
Further, the frequencies and the standard deviations that are reported in brackets received a special treatment, in order be read out more smoothly. 
The percent sign was eliminated from the header and incorporated into the numbers in brackets, which are now listed with a comma following the preceding values. 
Like any other numbered result reported in brackets with a descriptor in the header, the bracketed standard deviations is now stored in a seperate column. 
The row-wise collapsed content of the orginal table in Table \ref{tab:complexTable1} that is returned by \textit{table2text()} is shown in Table \ref{tab:complexTable1result}. 

\begin{table}[htbp]
\caption{Collapsed and unified content of the original table structure within Table \ref{tab:complexTable1}}
\begin{tabular}{l}
\toprule
\textit{table2text(x)}\\
\midrule
$\left[1\right]$ `Variable: Age, Category=20-25, Frequency=78, 39.79\%, Mean Income=1345, SD=250'\\
$\left[2\right]$ `Variable: Age, Category=26-31, Frequency=65, 33.16\%', Mean Income=1825, SD=535\\
$\left[3\right]$ `Variable: Age, Category=32 and older, Frequency=53, 27.05\%, Mean Income=2315, SD=930'\\
$\left[4\right]$ `Variable: Gender, Category: Men, Frequency=97, 49.49\%, Mean Income=1965, SD=550'\\
$\left[5\right]$ `Variable: Gender, Category: Women, Frequency=99, 50.51\%, Mean Income=1685, SD=350'\\
\midrule

\end{tabular}
\label{tab:complexTable1result}
\end{table}

\subsubsection{Example 2}
Table \ref{tab:complexTable2} represents an example of a much more complex multi-class table structure that reports descriptive measures and empirical correlations. 
For this table, it is most adequate to apply a mixture of the matrix- and the row-wise collapsing approach, with several special preperation issues. 
Information about how to decode the numbers in brackets and the asterisks is supplied in the footnotes. 
The table header only contains codes for the variable names, which are (partly) fully written out in the first column. 
Several correlations are reported with an asterisk as a code for p-values, others do not have such a code. 
Therefore, the absence of an asterisk is information as well, but only within the correlation matrix, and except for the values with r=1.

\begin{table}[htbp]
\caption{Example of a more complex multi-class table displaying descriptive measures and empirical correlations and its unified and preprocessed version}

\begin{minipage}{1\textwidth}
\centering
\begin{threeparttable}
\caption*{\textbf{Original table with table notes}}
\begin{tabular}{|l|c|c|c|}
\hline
& 1. & 2. & 3.\\
\hline
Mean & 45 (11.2) & 5,200 (21,123) & 26.8 (5.2)\\
\hline
Median & 40 & 3,200 & 25 \\ 
\hline
1. Age & 1 & .38** & .04\\
\hline
2. Income &  & 1 & .21*\\
\hline
3. Extr. & & & .87\\
\hline
\end{tabular}
\begin{tablenotes}
\item[N=125, $^*$ p<.05, $^{**}$ p<.01] 
\item[Numbers in brackets represent standard deviations.]
\item[Extr.: Extraversion]
\end{tablenotes}
\end{threeparttable}
\end{minipage}%

\bigskip

\begin{minipage}{1\textwidth}
\centering
\begin{threeparttable}
\caption*{\textbf{Unified and preprocessed matrix}}
\begin{tabular}{|l|c|c|c|}
\hline
& Age & Income & Extraversion \\
\hline
Mean & 45, SD=11.2 & 5200, SD=21123 & 26.8, SD=5.2\\
\hline
Median & 40 & 3200 & 25 \\ 
\hline
 Age & 1 & .38;; p<.01 & .04;; p>.05\\
\hline
 Income &  & 1 & .21;; p<.05\\
\hline
 Extraversion & & & .87\\
\hline
\end{tabular}
\end{threeparttable}
\end{minipage}

\label{tab:complexTable2}
\end{table}

The unified and preprocessed original table illustrates how the arguments to control the decoding of p-values (`\textit{decodeP}' and `\textit{noSign2P}'), extend abbreviations (`\textit{extendAbbreviations}'), and the special handling of correlation matrices work in practice. 
The coded p-values have been replaced with the seperator `;;' and the referenced value from the footnote. 
Only for the part of the correlation matrix, the correlations without an asterisk and an absolut value not equal to 1, a p-value greater the maximum of the coded p-values is attached. 

Table \ref{tab:complexTable2:result} contains the extended and unified input table, before it is collapsed with part specific rules (row-wise for the descriptives, matrix.wise for the correlation matrix). 
To give a first example of the special handling of degrees of freedom the argument `\textit{dfHandling}' is activated here. 
The reported N from the footnote is subtracted with 2 and added behind the letter r within brackets. 
Thanks to the imputation of N-2 to to correlations, the output can now be checked for consistency with \textit{standardStats()}. 

\begin{table}[htbp]
\caption{Collapsed and unified content of the original table structure within Table \ref{tab:complexTable2}}

\begin{tabular}{l}
\toprule
\textit{table2text(x, decodeP=TRUE, noSign2p=TRUE, expandAbbreviations=TRUE, dfHandling=TRUE)}\\
\midrule

$\left[1\right]$ `Mean, Age=45;; SD=11.2, Income=5200;; SD=21123, Extraversion=26.8;; SD=5.2'\\
$\left[2\right]$ `Median, Age=40, Income=3200, Extraversion=25'\\
$\left[3\right]$ `Age $<<\sim>>$ Age: r(123)=1'\\  
$\left[4\right]$ `Age $<<\sim>>$ Income: r(123)=.38;; p<.01'\\
$\left[5\right]$ `Age $<<\sim>>$ Extraversion: r(123)=.04;; p>0.05'\\
$\left[6\right]$ `Income $<<\sim>>$ Income: r(123)=1'\\
$\left[7\right]$ `Income $<<\sim>>$ Extraversion: r(123)=.21;; p<.05'\\
$\left[8\right]$ `Extraversion $<<\sim>>$ Extraversion: alpha=.87'\\

\midrule
\end{tabular}
\label{tab:complexTable2:result}
\end{table}

\pagebreak

\subsubsection{Example 3}
The next example table in Table \ref{tab:complexTable3} contains parameter estimates and model statistics for three different models. 
Again, it is reasonable to divide the table into parts and indivdually preprocess them. 
As an example for the processing of multi-line headers and connected cells, the table contains a two row-header with three connected cells in the first row. 
The standard errors of the coefficients are reported within brackets and the p-values are reported by simple codes. 

\begin{table}[htbp]
\caption{Example of a more complex table with a multi-row header displaying model parameters and statistics}

\begin{minipage}{1\textwidth}
\begin{center}

\begin{threeparttable}
\caption*{\textbf{Original table with table notes}}
\begin{tabular}{|l|c|c|c|}
\hline
 & \multicolumn{3}{c|}{Job performance}  \\
Variables & \multicolumn{1}{c}{Model 1} & \multicolumn{1}{c}{Model 2} & \multicolumn{1}{c|}{Model 3} \\
\hline
Income   & -0.26 (.16) & -0.39 (.21) & -0.36 (.18)\\
Extraversion   &  & 0.12\textsuperscript{+} (.06) & 0.16*** (.03)\\
Income * Extraversion   &  &  & -0.18** (.1)\\
\hline
R\textsuperscript{2} & 0.07 & 0.22 & 0.31\\
$\Delta$R\textsuperscript{2}  & 0.07 & 0.15 & 0.09\\
$\Delta$F  & 2.16 & 21.76** & 14.37*\\
\hline
\end{tabular}
\begin{tablenotes}
\item[+] p=.075; $^*$ p<.05; $^{**}$ p<.01; $^{***}$ p<.001
\item[] Numbers in brackets represent standard errors.
\end{tablenotes}
\end{threeparttable}\\
\vspace{.2in}
%\end{minipage}%
%\begin{minipage}{.5\textwidth}
\begin{threeparttable}
\caption*{\textbf{Unified and preprocessed matrix}}
\begin{tabular}{|l|c|c|c|}
\hline
 & Job performance& Job performance& Job performance  \\
 & Model 1 & Model 2 & Model 3 \\
\hline
Variables: Income   & -0.26, SE=.16& -0.39, SE=.21 & -0.36;; p>.05, SE=.18\\
\hline
Variables: Extraversion   &  & 0.12;; p=.075, SE=.06 & 0.16;; p<.001, SE=.03\\
\hline
Variables: Income * Extraversion   &  &  & -0.18*;; p<.05, SE=.1\\
\hline
 & R2=0.07, $\Delta$R2=0.07, & R2=0.22, $\Delta$R2=0.15, & R2=0.31, $\Delta$R2=0.09,\\
  & $\Delta$F=2.16;;p>0.05 & $\Delta$F=21.76:: p<0.01 & $\Delta$F=14.37;; p<.05\\
\hline
\end{tabular}
\end{threeparttable}
\end{center}
\end{minipage}

\label{tab:complexTable3}
\end{table}

The uniformization and preprocessing of the two table parts results in a matrix that can be collapsed adequately by the row-wise approach. 
The multi-row header structure was collapsed into a single row
The numbers in brackets and the p-values were decoded within the cells. 
The special handling of model result tables collapsed the last three rows into a single row after adding the statistic with an equal sign to the cells with numeric content in the same row. 
The distinction between the upper and the lower part of the table enables a nearly perfect imputation of `$p>\alpha$' in the absence of a p-value coding. 
The upper part is screened column-wise for coded p-values, and the lower part row-wise. 
If a column/row contains at least one coded p-value, all cells without a code are treated as nonsignificant. 
Here, this approach leads to falsely not imputed nonsignificant p-values. 
In the example table, the coefficients of Model 1 and Model 2 do not have a code for a significant p-value, and therefore, they are not recognized as candidate cells for imputation. 
In the lower part of the table, a nonsignificant p-value was correctly imputed to the cell without a code within the line with coded p-values.

\begin{table}[htbp]
\caption{Collapsed and unified content of the original table structure within Table \ref{tab:complexTable3}}
\begin{tabular}{l}
\toprule
\textit{table2text(x, decodeP=TRUE, noSign2p=TRUE, bracketHandling=TRUE, unifyStats=TRUE)}\\
\midrule
$\left[1\right]$ `Variables: Income, Job performance Model 1 b=-0.26, SE=0.16, Job performance Model 2 b=-0.39, \\\hspace{.25in}SE=0.21, Job performance Model 3 b=-0.36;; p>0.05, SE=0.18'\\
$\left[2\right]$ `Variables: Extraversion, Job performance Model 2 b=0.12;; p=0.075, SE=0.6, \\\hspace{.25in}Job performance Model 3 b=0.16;; p<0.001, SE=0.03'\\
$\left[3\right]$ `Variables: Income * Extraversion, Job performance Model 3 b=-0.18;; p<0.01, SE=0.1'\\
$\left[4\right]$ `Job performance Model 1: R2=0.07, delta R2=0.07, delta F=2.16;; p>0.05, \\\hspace{.25in}Job performance Model 2: R2=0.22, delta R2=0.15, delta F=21.76;; p<0.01, \\\hspace{.25in}Job performance Model 3: R2=0.31, delta R2=0.9, delta F=14.37;; p<0.05'\\
\midrule
\end{tabular}
\label{tab:complexTable3:result}
\end{table}

Table \ref{tab:complexTable3:result} contains the collapsed table, that was decoded with \textit{table2text()}, with several special handling options activated. 
Besides the already introduced options to handle coded p-values and numbers in brackets, the `\textit{unifyStats}' argument is activated. 
This is especially useful for the later extraction of statistical standard results, because they will be unified to a well handable format. 
Although nowhere mentioned within the table, the character `b' is added in front of every number that is followed by a standard error and p-value. 
Assuming that b/SE is a standard normal distribution, \textit{JATSdecoder}'s function \textit{standardStats()} can compute Z-values and recalculate p-values when the the equivalent \textit{table2text()} option `\textit{estimateZ}' is activated. 
This enables a consistency check for the p-values of the coefficients.

\subsubsection{Example 4}
Table \ref{tab:complexTable4} gives a last example with a simple ANOVA table. 
Along the column with p-values, bold and italic text styles are used to encode specific thresholds of p, that are reported in the table's footnote. 
In order to later check the reported F-statistic and p-value for consistency, it is reasonable to divide the table into two parts. 
F-values are accompanied by two different degrees of freedom values. 
Whereas the lines in the upper part of the table contain the factor specific degrees of freedom (df1), the lower part contains the residual (df2) and total degrees of freedom. 
The latter is irrelevant for a recomputation of the p-value. 

\begin{table}[htbp]
\caption{Example of a more complex table displaying an ANOVA result with style coded p-values}

\begin{minipage}[t]{1\textwidth}

\centering
\begin{threeparttable}
\caption*{\textbf{Original table with table notes}}

\begin{tabular}{lccccc}
\toprule
\textbf{Variable} & \textbf{SSq} & \textbf{df} & \textbf{MSq} & \textbf{F} & \textbf{P(>F)}\\
\midrule
Factor A & 12 & 2 & 6 & \textbf{9.09} & \textbf{0.00}\\
Factor B & 4.5 & 1 & 4.5 & \textit{6.82} & \textit{0.01}\\
Factor A*B & 3 & 2 & 1.5 & 2.27 & .12\\
\midrule
Residuals & 20 & 30 & 0.66 & & \\
Total& 39.5 & 35 & 1.13 & & \\
\midrule
\end{tabular}
\begin{tablenotes}
\item[Italic values are p<.05.]
\item[Bold valus indicate significance with p<.01.]
\end{tablenotes}
\end{threeparttable}
\end{minipage}%
\vspace{.5cm}
\begin{minipage}[t]{1\textwidth}
\centering
\begin{threeparttable}
\caption*{\textbf{Unified and preprocessed matrix}}
\begin{tabular}{|l|c|c|c|c|c|}
\hline
Variable & SSq & df & MSq & F & P(>F)\\
\hline
Factor A & 12 & 2 & 6 & 9.09;; p<.01 & 0.00;; p<.01\\
\hline
Factor B & 4.5 & 1 & 4.5 & 6.82;; p<.05 & 0.01;; p<.05\\
\hline
Factor A*B & 3 & 2 & 1.5 & 2.27;; p>.05 & .12;; p>.05\\
\hline
Residuals & 20 & 30 & 0.66 & & \\
\hline
Total& 39.5 & 35 & 1.13 & & \\
\hline
\end{tabular}
\end{threeparttable}
\end{minipage}

\label{tab:complexTable4}
\end{table}

The unified and preprocessed matrix shows the matrix before collapsing. 
The text styled numbers were transformed to significant and nonsignificant p-values, since both arguments, `\textit{decodeP}' and `\textit{noSign2p}', are activated. 
By activating the arguments `\textit{unifyStats}' and `\textit{dfHandling}' an invisible processing of the collapsed content is initiated. 
Here, the uniformization primarily involves the textual repesentation of the p-value and the degrees of freedom. 
Since the F- and the p-values are text styled to encode a p-threshold, only the last appearance is kept. 
The residual degrees of freedom in the lower part are detected and added to the lines with only factor specific degrees of freedom as `df2'. 
The factor specific degrees of freedom is renamed to `df1'. 
Now, each of the first three lines of the resulting text vector contains one checkable result by \textit{standardStats()}.

\begin{table}[htbp]
\caption{Collapsed and unified content of the original table structure within Table  \ref{tab:complexTable4}}
\begin{tabular}{l}
\toprule
\textit{table2text(x, decodeP =TRUE, noSign2p = TRUE, unifyStats = TRUE, dfHandling = TRUE)}\\
\midrule
$\left[1 \right]$ `Variable: Factor A, SSq=12, df1=2, df2=30, MSq=6, F=9.09, p=0.00;; p<0.01' \\
$\left[2 \right]$ `Variable: Factor B, SSq=4.5, df1=1, df2=30, MSq=4.5, F=6.82, p=0.01;; p<0.05'\\
$\left[3 \right]$ `Variable: Factor A * B, SSq=3, df1=2, df2=30, MSq=1.5, F=2.27, p=0.12;; p>0.05'\\
$\left[4 \right]$ `Variable: Residuals, SSq=20, df2=30, MSq=0.66' \\
$\left[5 \right]$ `Variable: Total, SSq=39.5, df=35, MSq=1.13' \\
\midrule
\end{tabular}
\label{tab:complexTable4:result}
\end{table}

\pagebreak
\subsection{Extraction and consistency checkup of statistical standard results}

The \textit{get.stats()} function of the \textit{JATSdecoder} package is a reliable detector for statistical results within textual reports \cite{getStats}. 
Its function \textit{allStats()} returns all numerical results within the full text of a manuscript as a vector with sticked results. 
Sticked results are represented by any sequence of letter-number combinations pointing to numbers with an operator (e.g., test statistic, degrees of freedom, p-value). 
The sticked results are then screened for statistical standard results (t-, Z-, F-, $\chi^2$-, $\beta$-, r-, $R^2$-, U-, H-, Q-, $G^2$-, d-, $\eta^2$-, $\omega^2$, OR, RR, and p-values, as well as Bayes Factors, degrees of freedom, and standard errors) with the \textit{standardStats()} function. 
The resulting data frame only contains those of the sticked results that contain statistical standard results. 
If a result is reported with sufficient information, the p-value is recomputed and can be compared to the reported p-value.

The format of the decoded table content created with \textit{table2text()} is well suited to be processed with the \textit{standardStats()} function. 
Quick access with several user-adjustable options and table-specific preprocessing is enabled with \textit{tableParser}'s function \textit{table2stats()}. 
In contrast to applying \textit{standardStats()} on the raw collapsed text, \textit{table2stats()} decodes the tabled content with all introduced settings activated. 
Since \textit{standardStats()} expects a vector with only one sticked result per cell, lines with several test results are split up at adequate positions. 
The option to decode p-values with the common asterisk-based standard coding scheme via activating the argument `\textit{standardPcoding}' enables a detection and consistency check for test results, even when no decoding scheme is reported or extracted from the footnote.

By activating the option `\textit{checkP}', an automated consistency checkup of the reported and/or coded p-values with the recomputed p-values may be initiated. 
The checkup can return specific types of inconsistency (false positive/negative p-value/coded p-value, numeric inconsistency in p-value/coded p-value).
In standard mode, the critical threshold for differences in the reported and recalculated p-values is .02, with a sharp border at the $\alpha$ level. 
The value of $\alpha$ is essential to differentiate between a numeric and a decisive inconsistency. 
For example, a reported p-value of .0501 is declared as a false negative when the recalculated p-value is 0.0499 and $\alpha=.05$ is applied. 
A reported p-value of .01 is declared as a numerical inconsistency when the recaclculated p-value is between .03 and .05, and as false positive when the recalculated p-value is above or equal to .05. 
P-values for one-sided tests can be calculated by setting the argument `\textit{alternative}' to `directed'. 
Thus, the direction of the statistical conclusion has to be checked manually. 

The $\alpha$ level can be globally set for all checks within a document with the argument `\textit{alpha}'. 
In standard mode, \textit{table2stats()} tries to identify a table-specific $\alpha$ level from within the footnotes with \textit{JATSdecoder}'s function \textit{get.alpha.error()}. 
The fallback value of the $\alpha$ level detector is the common standard of $\alpha=.05$. 

Generally, \textit{table2stats()} decodes sign-coded p-values into a textual representation. 
Due to its rather low accuracy, the imputation of p-values within cells that do not have a coded p-value is deactivated in standard mode. 
It can be initiated via the argument `\textit{noSign2p}'. 

In statistics, the t-distribution converges to the standard normal distribution with increasing sample size.
Assuming a sufficiently large sample, t-statistics can be interpreted as z-values. 
To increase the possibility for recalculating and checking p-values the argument `\textit{estimateZ}' can be activated. 
For test results that contain an effect measure (`d' or `b') and a standard error but no fully reported test statistic (t-value with degrees of freedom, z-value), an estimate for z (`d/SE' or `b/SE') and the corresponding p-value is calculated. 
If a result contains a t-value but no value for the degrees of freedom, then t is treated as an estimate for z, and the p-value is computed. 

Lastly, \textit{JATSdecoder}'s function \textit{get.multi.comparison()} is applied to the footnotes. 
It detects references to statistical methods that control for cumulating $\alpha$ errors in scenarios with multiple statistical tests. 
Since the cumulation of $\alpha$ errors may be compensated in several ways (corrected p-values or corrected $\alpha$ level), a warning message concerning the adequateness of the performed checks is returned if a correction method for multiple testing is detected. 
For a clear identification inside the data frame output, the table-specific name of the correction method is attached to every checked result. 
 
The final output can be returned as a list of data frames with the extracted standard results per table, or as a single data frame object for all tables within a document. 
The data frames with the detected standard results within the complex table examples 2--4 are presented in Table \ref{tab:stats1}--\ref{tab:stats3}. 
They were created with a single command on the example DOCX file path (x) with:\\

\textit{table2stats(x, checkP=TRUE, estimateZ=TRUE, noSign2p=TRUE, collapse=FALSE)}\\ 

The results are returned table-wise, since the argument `\textit{collapse}' is deactivated. 
The output does not contain the collapsed text of Table \ref{tab:simpleTable} and Table \ref{tab:complexTable1}, since they do not contain any statistical standard results. 
The last columns with the $\alpha$ level used for the check and the detected correction procedure are ommited in Table \ref{tab:stats2} and Table \ref{tab:stats3} to reduce the table width. 
For the same reason, the empty column containing the detected error type is omitted from Table \ref{tab:stats1} and Table \ref{tab:stats3}. 

\begin{landscape}

\begin{table}[ht]
\centering
\small
\addtolength{\tabcolsep}{-0.2em}
\caption{Data frame with the extracted and checked statistical results within example Table \ref{tab:complexTable2}}
\begin{tabular}{lccccccccccc}
  \hline
result & r\_op & r & df2 & codedP\_op & codedP & recalculatedP & deltaP2tailed & error & alpha4check & correction\_meth \\ 
  \hline
Gender $<<\sim>>$ Gender: r(123)=1 & = & 1 & 123 &  &  & 0 &  &    \\ 
Gender $<<\sim>>$ Income: r(123)=0.38;; p$<$0.01 & = & 0.38 & 123 & $<$ & 0.01 & 0 & -0.01 & 0 & 0.05 & Bonferroni, Holm \\ 
Gender $<<\sim>>$ Extraversion: r(123)=0.04;; p$>$0.05 & = & 0.04 & 123 & $>$ & 0.05 & 0.6578 & 0.6078 & 0 & 0.05 & Bonferroni, Holm\\ 
Income $<<\sim>>$ Income: r(123)=1 & = & 1 & 123 &  &  & 0 &  &    \\ 
Income $<<\sim>>$ Extraversion: r(123)=0.21;; p$<$0.05 & = & 0.21 & 123 & $<$ & 0.05 & 0.0187 & -0.0313 & 0  & 0.05 & Bonferroni, Holm\\ 
   \hline
\end{tabular}
\label{tab:stats1}
\end{table}

\begin{table}[ht]
\centering
\footnotesize
\addtolength{\tabcolsep}{-0.4em}
\caption{Data frame with the extracted and checked statistical results within example Table \ref{tab:complexTable3}}

\begin{tabular}{lcccccccccccc}
  \midrule
 result & beta & SEbeta & Zest & R2\_op & R2 & codedP\_op & codedP & recalc.P & deltaP2t. & error & errorType \\ 
  \midrule
 Variables: Income, Job performance Model 1 b=-0.26, SE=0.16 & -0.26 & 0.16 & -1.625 &  &  &  &  & 0.1042 &  &  &  \\ 
 Variables: Income, Job performance Model 2 b=-0.39, SE=0.21 & -0.39 & 0.21 & -1.857 &  &  &  &  & 0.0633 &  &  &  \\ 
 Variables: Income, Job performance Model 3 b=-0.36;; p$<$0.01, SE=0.1 & -0.36 & 0.1 & -3.6 &  &  & $<$ & 0.01 & 3e-04 & -0.0097 & 0 &  \\ 
 Variables: Extraversion, Job performance Model 2 b=0.12;; p=0.075, SE=0.6 & 0.12 & 0.6 & 0.2 &  &  & = & 0.075 & 0.8415 & 0.7665 & 1 & num. inc. in p coding \\ 
 Variables: Extraversion, Job performance Model 3 b=0.16;; p$<$0.001, SE=0.03 & 0.16 & 0.03 & 5.333 &  &  & $<$ & 0.001 & 0 & -0.001 & 0 &  \\ 
 Variables: Income * Extraversion, Job performance Model 3 b=-0.18;; p$<$0.01, SE=0.1 & -0.18 & 0.1 & -1.8 &  &  & $<$ & 0.01 & 0.0719 & 0.0619 & 1 & false positive p coding \\ 
 Job performance Model 1: R2=0.07, delta R2=0.07, delta F=2.16;; p$>$0.05 &  &  &  & = & 0.07 & $>$ & 0.05 &  &  &  &  \\ 
Job performance Model 2: R2=0.22, delta R2=0.15, delta F=21.76;; p$<$0.01 &  &  &  & = & 0.22 & $<$ & 0.01 &  &  &  &  \\ 
Job performance Model 3: R2=0.31, delta R2=0.9, delta F=14.37;; p$<$0.05 &  &  &  & = & 0.31 & $<$ & 0.05 &  &  &  &  \\ 
   \midrule
\end{tabular}
\label{tab:stats2}
\end{table}

\begin{table}[ht]
\centering
\small
\addtolength{\tabcolsep}{-0.4em}
\caption{Data frame with the extracted and checked statistical results within example Table \ref{tab:complexTable4}}
\begin{tabular}{lccccccccccc}
  \toprule
result & F\_op & F & df1 & df2 & p\_op & p & codedP\_op & codedP & recalc.P & deltaP2tailed & error  \\ 
  \midrule
Variable: Factor A, SSq=12, df1=2, df2=30, MSq=3, F=9.09, p=0.00;; p<0.01 & = & 9.09 & 2 & 30 & = & 0 & $<$ & 0.01 & 0.0008  & 0.0008 & 0 \\ 
Variable: Factor B, SSq=4.5, df1=1, df2=30, MSq=4.5, F=6.82, p=0.01;; p<0.05 & = & 6.82 & 1 & 30 & = & 0.01 & $<$ & 0.05 & 0.0139 & 0.0039  & 0 \\ 
Variable: Factor A * B, SSq=3, df1=2, df2=30, MSq=1.5, F=2.27, p=0.12;; p>0.05 & = & 2.27 & 2 & 30 & = & 0.12 & $>$ & 0.05 & 0.1208 & 0.0008 & 0 \\ 
   \midrule
\end{tabular}
\label{tab:stats3}
\end{table}

\end{landscape}

All five reported correlations and the coded p-values in Table \ref{tab:complexTable2} are detected and correctly processed. 
Since values of `r=1' should not be tested, only three of the five correlations are checked for consistency. 
Using the standard threshold for critical differences in the reported and recalculated p-value of .02 and the standard $\alpha$ level of 5\%, no obvious inconsistency is detected. 
However, the footnote cites the use of the Bonferroni-Holm correction procedure, and the recalculated p-values are not corrected. 
A check with the nominal $\alpha$ level is therefore not appropriate, if uncorrected p-values are reported. 
Further, corrected p-values should not be compared to uncorrected p-values. 

Since the results in Table \ref{tab:complexTable3} do not contain a test statistic that would enable a recalculation of the p-values, the estimation of z-values by dividing the beta coefficients by the standard errors enables a consistency check for all tested coefficients. 
The reported model statistics do not contain the necessary degrees of freedoms to recalculate the p-values, and therefore, no checks can be performed. 
The consistency check detects two errors. 
The first one is tagged as numeric inconsistency, since the recalculated p-value does not affect the statistical decision. 
The second inconsistency is tagged as a decisive error because the coded and recalculated p-values lead to opposing statistical decisions. 

The consistency check for the three tested factors within the ANOVA table in Table \ref{tab:complexTable4} is successful and shows no inconsistency for the reported and coded p-values when compared to the recalculated p-values with a critical difference of .02. 
The check is only possible with the special handling of degrees of freedom, which is activated in standard mode in \textit{table2stats()}.

\begin{comment}
\subsection{Caption and footer extraction}

The functions \textit{get.caption()} and \textit{get.footer()} can be applied to tables coded in NISO-JATS format. The functions return the detected caption/footer text in native HTML or human readable format. 
Again, letter uniformisation can be initiated by activating the option `letter.convert. HTML tags can be removed with `\textit{rm.html}'.
Further, the activating the argument \textit{sentences} allows to output the text as sentence vector created with \textit{text2sentences()}. 

\end{comment}

\section{Conclusion}
\label{sec:conclusion}
The \textit{tableParser} functions enable a robust extraction and processing of simple and particular, more complex table structures. 
They can be used to store and process three-part-table structures in common file formats. 
However, table-to-matrix extraction is not feasible for TEX files, only an HTML render. 

The collapsed and decoded table as a text vector enables the interpretation of the tabled content without the table-specific navigation within rows and columns. 
Furthermore, the collapsed table structure is built up in a manner that allows the extraction of statistical standard results with \textit{JATSdecoder}'s function \textit{standardStats()}.
This expands the possibilities for extracting and post-processing statistical results in literature databases. 

The data frames with the reported statistical standard results created with \textit{table2stats()} open new possibilities for large scale investigations of the research literature and effects of specific variables. 
Since the content of correlation matrices is collapsed into a clearly identifiable format (`Variable A $<<\sim>>$ Variable B, r(df)=number'), a landscape or summary of the linear correlations of a specific variables with others can be drawn from the output of \textit{table2stats()}. 
The availablility of the reported degrees of freedom further enables to relate specific results to the size of the sample. 
For meta analyses, a search for specific covariates in table structures may increase the identifiability of relevant studies. 

Authors who are still in the publication process of an article with statistical test results presented in tables, may benefit from the tool as a complexity feedback and/or a consistency checkup engine for p-values.
It further motivates to produce barrier-free tables and may prevent from reporting inconsistant, or too strictly rounded results. 

Although large language models might be well suited to decode and review content in scientific tables, their probabilistic nature will very likely lead to slightly different results when applied multiple times to the same input. 
In contrast, the deterministic nature of the \textit{tableParser} functions guarantees reproducibility. 
Processing anomalies are easily traceable. 
Further, large language models are quite resource-consuming, especially when applied to large document collections. 
To create a database of tabled content within large document collections, \textit{tableParser}'s functions can be easily applied to a list of files with \textit{lapply()}, or, to shorten processing time, with \textit{future\_lapply()} from the \textit{future.apply} package \cite{future} for multi-core processing (e.g., \textit{`future\_lapply(vector.with.file.names,table2stats,check=TRUE)'}).

For large-scale analyses of research articles, a valuable open-access file resource with more than 7 million documents related to the medical and health sciences that were published by more than 23k journals is supplied by PubMed Central \cite{PMC}.
The full document collection can be bulk downloaded as NISO-JATS-encoded XML files at \url{https://ftp.ncbi.nlm.nih.gov/pub/pmc/oa_bulk/}.

\section{Limitations}
\label{sec:limitations}
The here listed example table structures within the example DOCX file were correctly extracted and adequately processed. 
However, in practice, table structures can be a lot more complex and may fail to render adequately with \textit{tableParser}. 

The table processing depends on technically clean coded tables. 
Although visually appearing like cells in several rows, content from within cells with line breaks is considered as content within one cell. 
The same problem arises with tabulatur-seperated cell content. 
For HTML and DOCX documents, the table extraction is only adequate for NISO-JATS-encoded three-part-table structures and the floating paragraphs approach. 
Captions and footnotes that are stored within a table's cell are treated as ordinary text cells. 

Several severe limitations are connected to the general technical problems with detecting and processing content stored within PDF documents. 
For any processing of PDF input, \textit{tableParser} depends on the conversion accuracy of the \textit{tabulapdf} package. 
The example PDF document (see appendix or GitHub repository) was created with \textit{LibreOffice} from the example DOCX document and illustrates this problem. 
Running the command \textit{table2matrix()} on the PDF version shows that the table-to-character matrix processing is erroneous for the example multi-model table (Table  \ref{tab:complexTable3}). 
The error is caused by the connected cells in the first header line. 
Text stylings, like those within the example ANOVA table, are never detected. 
Further, \textit{tabulapdf} will treat table structures that have a page break as seperate tables, although it is just a typesetting-caused split. 

For statistical result tables, the content decoding can often only be realized by relating the table's caption and/or footnote text to the table's content. 
Since neither, table captions nor footnotes, can be extracted from PDF documents \textit{tabulapdf}, the decoding depth is very limited. 
This also affects the decoding of encoded p-values and therefore lowers the checkability of results with \textit{standardStats()}.

The decision of the table classifier may introduce an inadequate collapsing rule (row- vs. matrix-wise). 
For table structures that contain a mixture of row- and matrix-based tables or a list of tables, it might be more appropriate to split the structure into a set of simple tables before collapsing. 
Standard results that were collapsed by the wrong approach lower the detection and checking accuracy. 
For example, when a correlation matrix is misinterpreted as an unspecific matrix-like table or simple tabled result, none of the numerical values can be identified as a correlation. 
Or when an ordinary result table is treated as a correlation matrix, this results in numerous falsely detected correlations.

%\section*{Author Biography}
%Ingmar B\"oschen is a lecturer of statistics and data analysis at the department of Psychological Methods and Statistics at the Institute of Psychology, University Hamburg. 

%\section*{ORCID}
%Ingmar Böschen; ORCID: 0000-0003-1159-3991

%%% REFERENCES
\bibliographystyle{apacite}

%\bibliography{/home/ingmar/JATSdecoderEvaluation/referencesAPA6bib.bib}   

\section*{Declarations}

\subsection*{Contributions}
Ingmar Böschen developed \textit{tableParser}, designed the example documents and wrote the manuscript.

\subsection*{Conflict of interest} 
The author declares no conflict of interest.

\subsection*{Funding}
No funding.

\subsection*{Availability of code and data}
\begin{itemize}
\item \textit{tableParser} and \textit{JATSdecoder} are freely available R packages that can be installed via CRAN or their GitHub accounts \url{https://github.com/ingmarboeschen/}.

\item The example DOCX file that was processed here, as well as an export as HTML and PDF can be accessed at \url{https://github.com/ingmarboeschen/tableParser}. Code examples are provided.

\item Single file processing can be realised via a document upload within the \textit{JATSdecoder} and \textit{tableParser} web interface \url{https://www.get-stats.app}
\end{itemize}

\end{document}